\documentclass[conference]{IEEEtran}
\IEEEoverridecommandlockouts
\usepackage{amsmath,amssymb,amsfonts}
\usepackage{algorithmic}
\usepackage{graphicx}
\usepackage{textcomp}
\usepackage{color}
\usepackage{xcolor}
\usepackage{hyperref}
\usepackage[caption=false,font=footnotesize]{subfig}
\usepackage{booktabs}
\usepackage{multirow}
\usepackage[comma, sort&compress,square,numbers]{natbib}
\usepackage{soul}
\usepackage{authblk}
\def\BibTeX{{\rm B\kern-.05em{\sc i\kern-.025em b}\kern-.08em
    T\kern-.1667em\lower.7ex\hbox{E}\kern-.125emX}}

\begin{document}

\title{BVI-AOM: A New Training Dataset for Deep Video Compression Optimization\\
\thanks{The authors would like to acknowledge funding from Netflix Inc., University of Bristol, and the UKRI MyWorld Strength in Places Programme (SIPF00006/1). For the purpose of open access, the author has applied a Creative Commons Attribution (CC BY) license to any Author Accepted Manuscript version arising.}
}

\author[1]{Jakub Nawała\IEEEauthorrefmark{1}\thanks{\noindent\IEEEauthorrefmark{1}Equal contribution.}}
\author[1]{Yuxuan Jiang\IEEEauthorrefmark{1}}
\author[1]{Fan Zhang}
\author[2]{Xiaoqing Zhu}
\author[2]{Joel Sole}
\author[1]{David Bull}
\affil[1]{\textit{Visual Information Laboratory, University of Bristol, Bristol, BS1 5DD, United Kingdom}}
\affil[1]{\textit {\{jakub.nawala, yuxuan.jiang, fan.zhang, dave.bull\}@bristol.ac.uk}}
\affil[2]{\textit{Netflix Inc., Los Gatos, CA, USA, 95032}}
\affil[2]{\textit {\{xzhu, jsole\}@netflix.com}}

\maketitle

\begin{abstract}

Deep learning is now playing an important role in enhancing the performance of conventional hybrid video codecs. These learning-based methods typically require diverse and representative training material for optimization in order to achieve model generalization and optimal coding performance. However, existing datasets either offer limited content variability or come with restricted licensing terms constraining their use to research purposes only. To address these issues, we propose a new training dataset, named BVI-AOM, which contains 956 uncompressed sequences at various resolutions from 270p to 2160p, covering a wide range of content and texture types. The dataset comes with more flexible licensing terms and offers competitive performance when used as a training set for optimizing deep video coding tools. The experimental results demonstrate that when used as a training set to optimize two popular network architectures for two different coding tools, the proposed dataset leads to additional bitrate savings of up to 0.29 and 2.98 percentage points in terms of PSNR-Y and VMAF, respectively, compared to an existing training dataset, BVI-DVC, which has been widely used for deep video coding. The BVI-AOM dataset is available at
 \url{https://github.com/fan-aaron-zhang/bvi-aom}.
\end{abstract}


\section{Introduction}

In recent years, the amount of video content sent over the Internet has increased significantly \cite{sandvine2023global}. Although an average Internet user has faster access to the network than before \cite{cisco2020annual}, the transmission throughput is still generally limited due to the increased user numbers and the more immersive video data consumed. In this context, video coding is now as important as ever. In the past twenty years, a series of video coding standards have been released by MPEG, including the most widely used video coding standard, H.264/Advanced Video Coding (AVC) \cite{h264AVC}, and its successors, H.265/High Efficiency Video Coding (HEVC) \cite{h265HEVC} and H.266/Versatile Video Coding, VVC \cite{h266VVC}. In the same vein, several video technology companies have formed a consortium, named Alliance of Open Media (AOM), aiming at developing open-source, royalty-free video coding standards, with its latest contribution, AOMedia Video 1 (AV1) \cite{chen2020overview}. 

\begin{figure}[!t]
    \centering
    \includegraphics[trim={0.4cm 3cm 0.35cm 3cm},clip,width=0.95\linewidth]{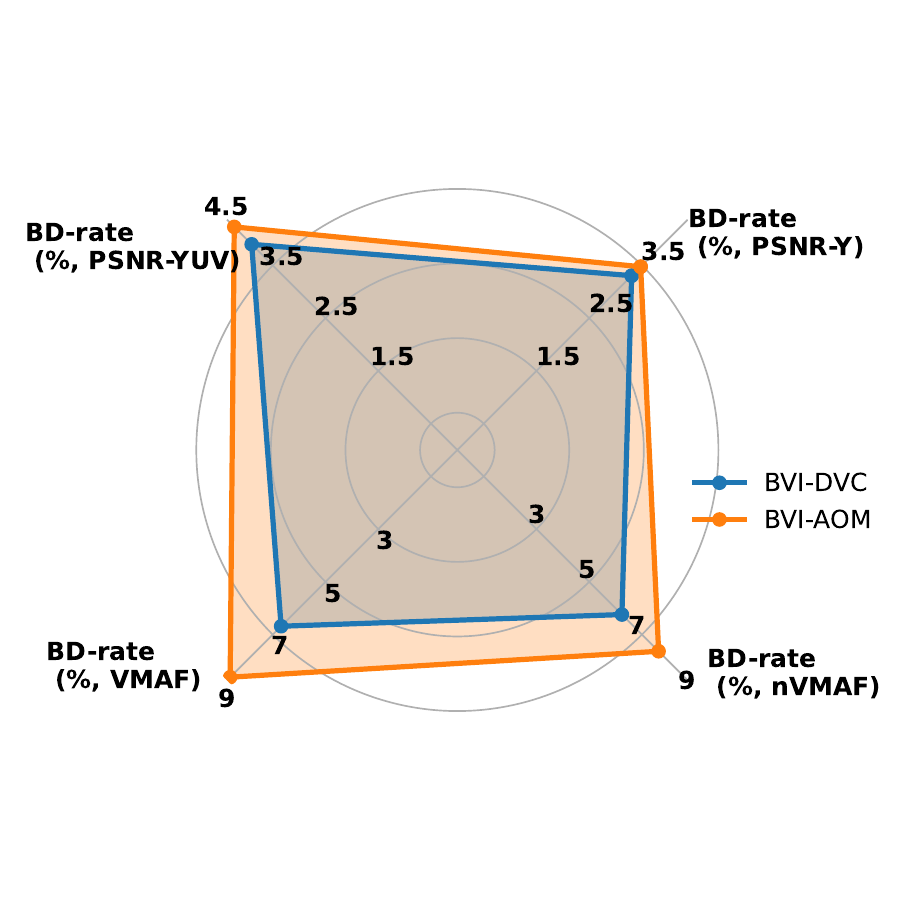}
    \vspace{-0.3cm}
    \caption{Radar chart comparing the average BD-rate savings over the anchor codec (AVM) for two different models and two coding tools,  trained either with the BVI-DVC or the BVI-AOM dataset, measured by four different quality metrics. Larger quadrilateral area indicates better performance.}
    \label{fig:radar-chart}
\vspace{-10pt}
\end{figure}

\begin{figure*}[t]
\begin{minipage}[b]{0.192\linewidth}
    \centering
    \centerline{\includegraphics[width=\linewidth]{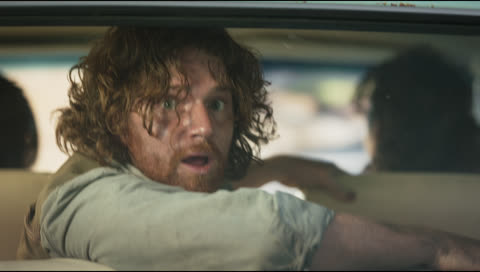}}
    \end{minipage}
    \begin{minipage}[b]{0.192\linewidth}
    \centering
    \centerline{\includegraphics[width=\linewidth]{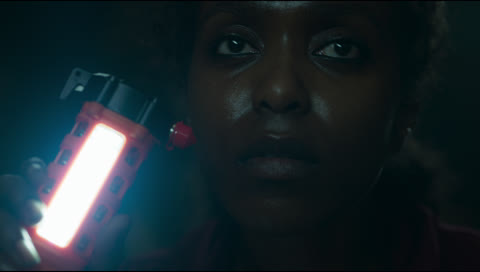}}
    \end{minipage}
    \begin{minipage}[b]{0.192\linewidth}
    \centering
    \centerline{\includegraphics[width=\linewidth]{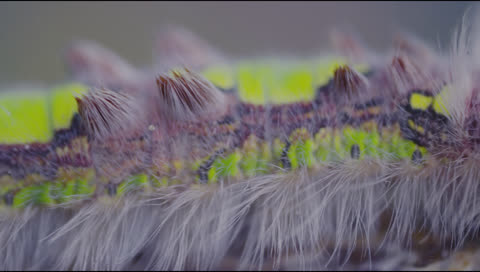}}
    \end{minipage}
    \begin{minipage}[b]{0.192\linewidth}
    \centering
    \centerline{\includegraphics[width=\linewidth]{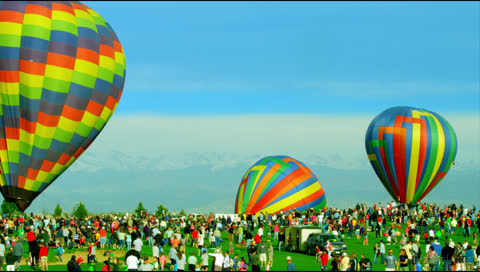}}
    \end{minipage}
    \begin{minipage}[b]{0.192\linewidth}
    \centering
    \centerline{\includegraphics[width=\linewidth]{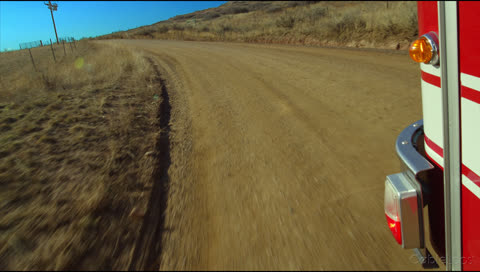}}
    \end{minipage}
    
    \begin{minipage}[b]{0.192\linewidth}
    \centering
    \centerline{\includegraphics[width=\linewidth]{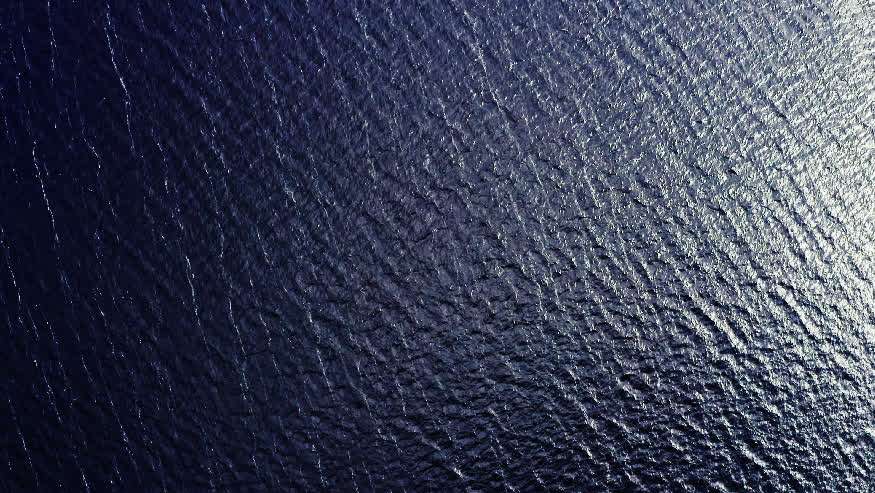}}
    \end{minipage}
    \begin{minipage}[b]{0.192\linewidth}
    \centering
    \centerline{\includegraphics[width=\linewidth]{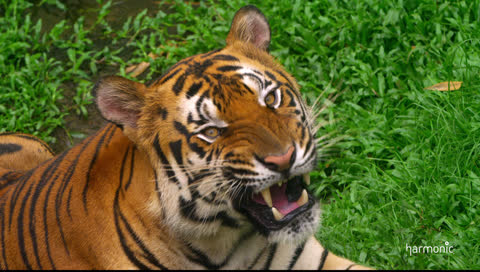}}
    \end{minipage}
    \begin{minipage}[b]{0.192\linewidth}
    \centering
    \centerline{\includegraphics[width=\linewidth]{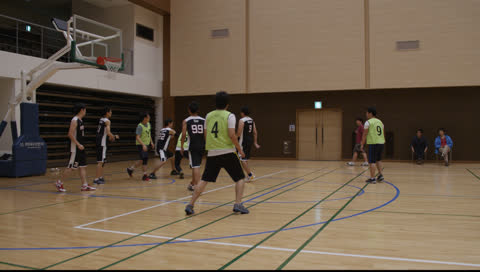}}
    \end{minipage}
    \begin{minipage}[b]{0.192\linewidth}
    \centering
    \centerline{\includegraphics[width=\linewidth]{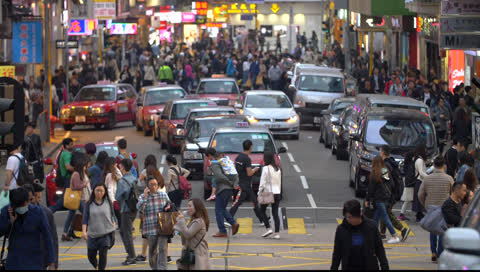}}
    \end{minipage}
    \begin{minipage}[b]{0.192\linewidth}
    \centering
    \centerline{\includegraphics[width=\linewidth]{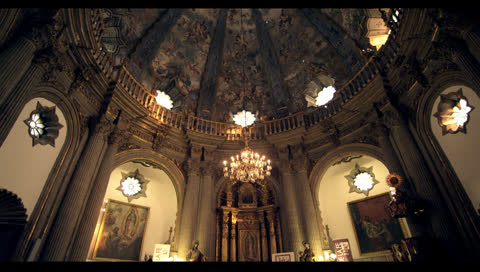}}
    \end{minipage}
    
    \begin{minipage}[b]{0.192\linewidth}
    \centering
    \centerline{\includegraphics[width=\linewidth]{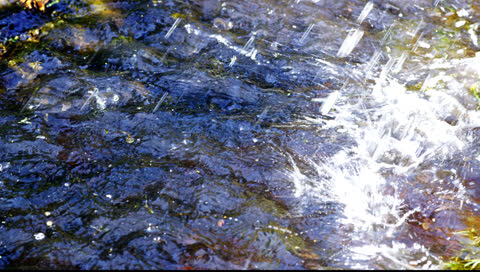}}
    \end{minipage}
    \begin{minipage}[b]{0.192\linewidth}
    \centering
    \centerline{\includegraphics[width=\linewidth]{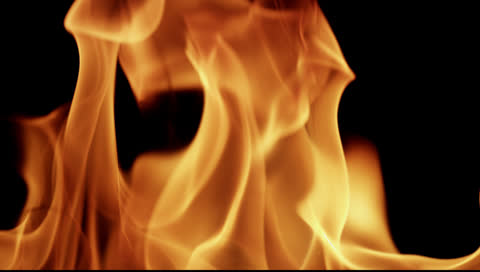}}
    \end{minipage}
    \begin{minipage}[b]{0.192\linewidth}
    \centering
    \centerline{\includegraphics[width=\linewidth]{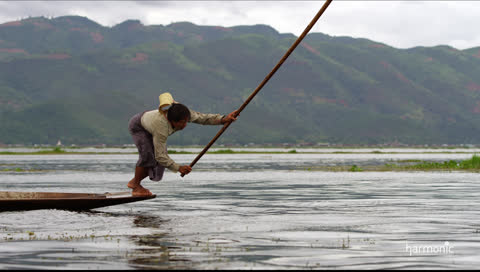}}
    \end{minipage}
    \begin{minipage}[b]{0.192\linewidth}
    \centering
    \centerline{\includegraphics[width=\linewidth]{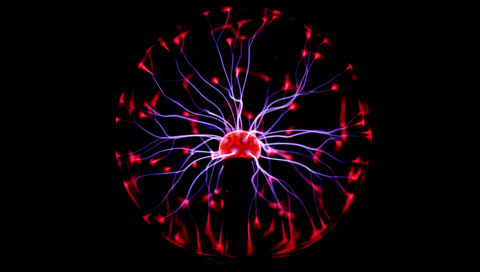}}
    \end{minipage}
    \begin{minipage}[b]{0.192\linewidth}
    \centering
    \centerline{\includegraphics[width=\linewidth]{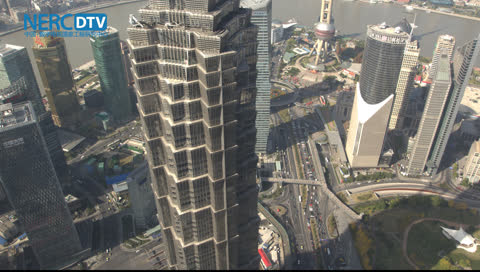}}
    \end{minipage}
    \caption{Thumbnails of 15 representative sequences from the BVI-AOM dataset.}
    \label{fig:thumbnails}
\end{figure*}

More recently, both MPEG and AOM have initialized new working models, in order to achieve further coding gains over the latest standards, H.266/VVC and AV1. In these models, in addition to sophisticated modifications to conventional coding modules, there are active investigations on the use of various deep learning techniques to obtain more evident improvement \cite{li2023designs, misra2023reduced, joshi2023switchable, zhang2020enhancing, zhang2021video, jia2023deep}. In parallel, deep neural networks have also been employed to build new coding frameworks \cite{kwan2023hinerv, dcvc, dcvcfm} which enable end-to-end optimization. Although these neural codecs typically require additional graphics processing resources and are associated with high computational complexity, they do show great potential to compete with standard video coding algorithms.

Most of these learning-based video codecs (except those based on implicit video representation models \cite{chen2023hnerv, kwan2023hinerv}), are commonly trained offline to obtain generic models before being inferred online for deployment. In these cases, the training content is essential for ensuring model generalization and optimal performance. To this end, several public video datasets have been developed specifically for deep video compression. One important example is BVI-DVC \cite{ma2022bvidvc}, which contains 800 video sequences with various spatial resolutions up to 2160p. Due to its content diversity and uniformity, it has been used by MPEG JVET for developing neural network-based coding tools. However, this dataset lacks certain content such as dark or high-contrast scenes, and its copyright license restricts its use in a wider community. 
Tencent Video Dataset (TVD) \cite{xu2021video} is another dataset developed for learning-based video compression, with 86 source sequences at the UHD resolution. Although it contains much fewer sequences compared to the BVI-DVC, the analysis performed in \cite{xu2021video} shows its relatively wide coverage in terms of encoding complexity. Other notable training datasets for deep video compression include DIV2K \cite{agustsson2017ntire}, Vimeo \cite{xue2019video}, REDS \cite{nah2019ntire} and HIF \cite{li2019deep}. 

\begin{figure}[t]
    \centering
    \begin{minipage}[b]{0.48\linewidth}
        \centering
        \centerline{\includegraphics[trim={1.9cm 0.5cm 1.8cm 0.5cm},clip,width=\linewidth]{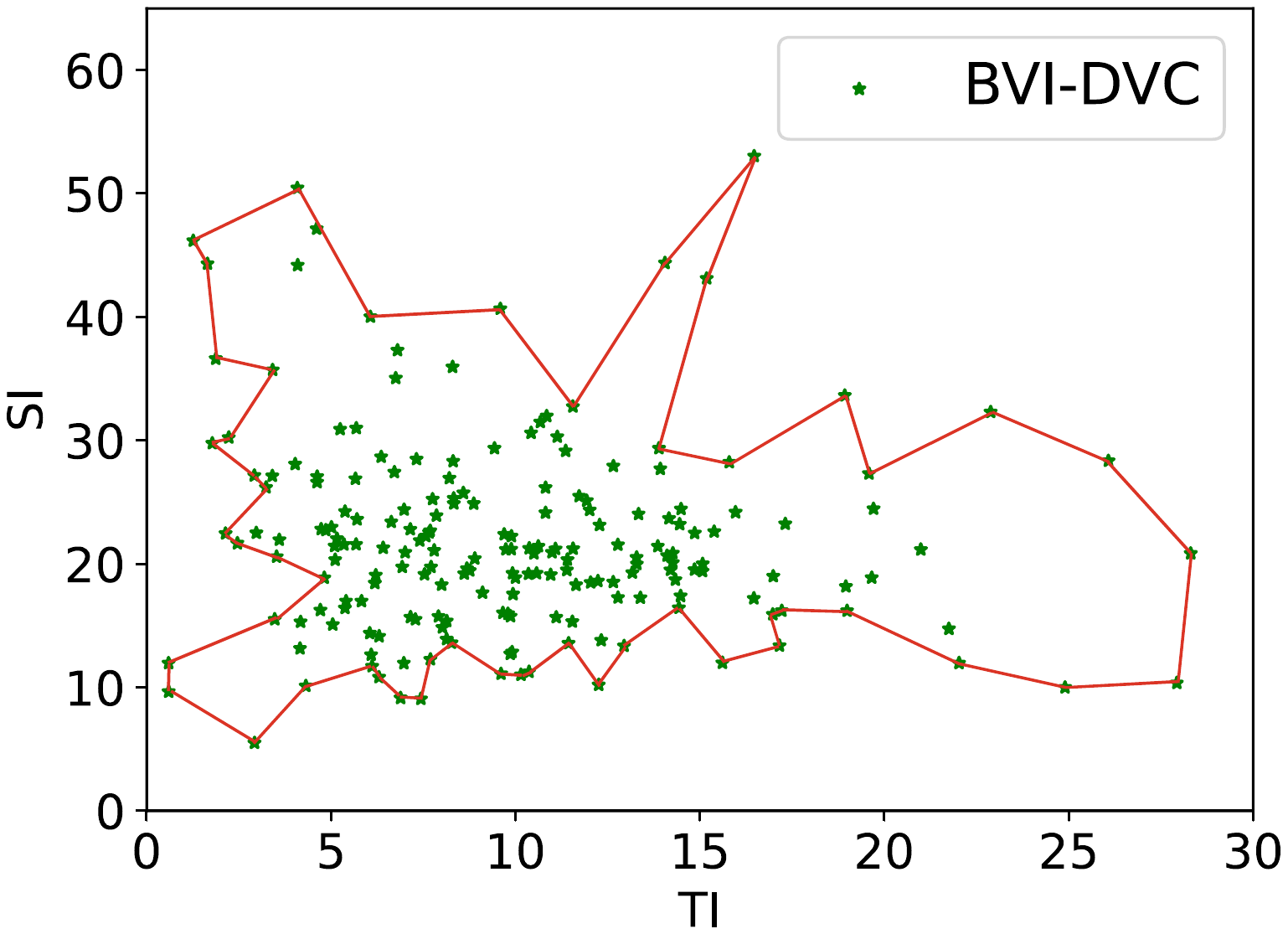}}
    \end{minipage}
    \begin{minipage}[b]{0.48\linewidth}
        \centering
        \centerline{\includegraphics[trim={1.9cm 0.5cm 1.8cm 0.5cm},clip,width=\linewidth]{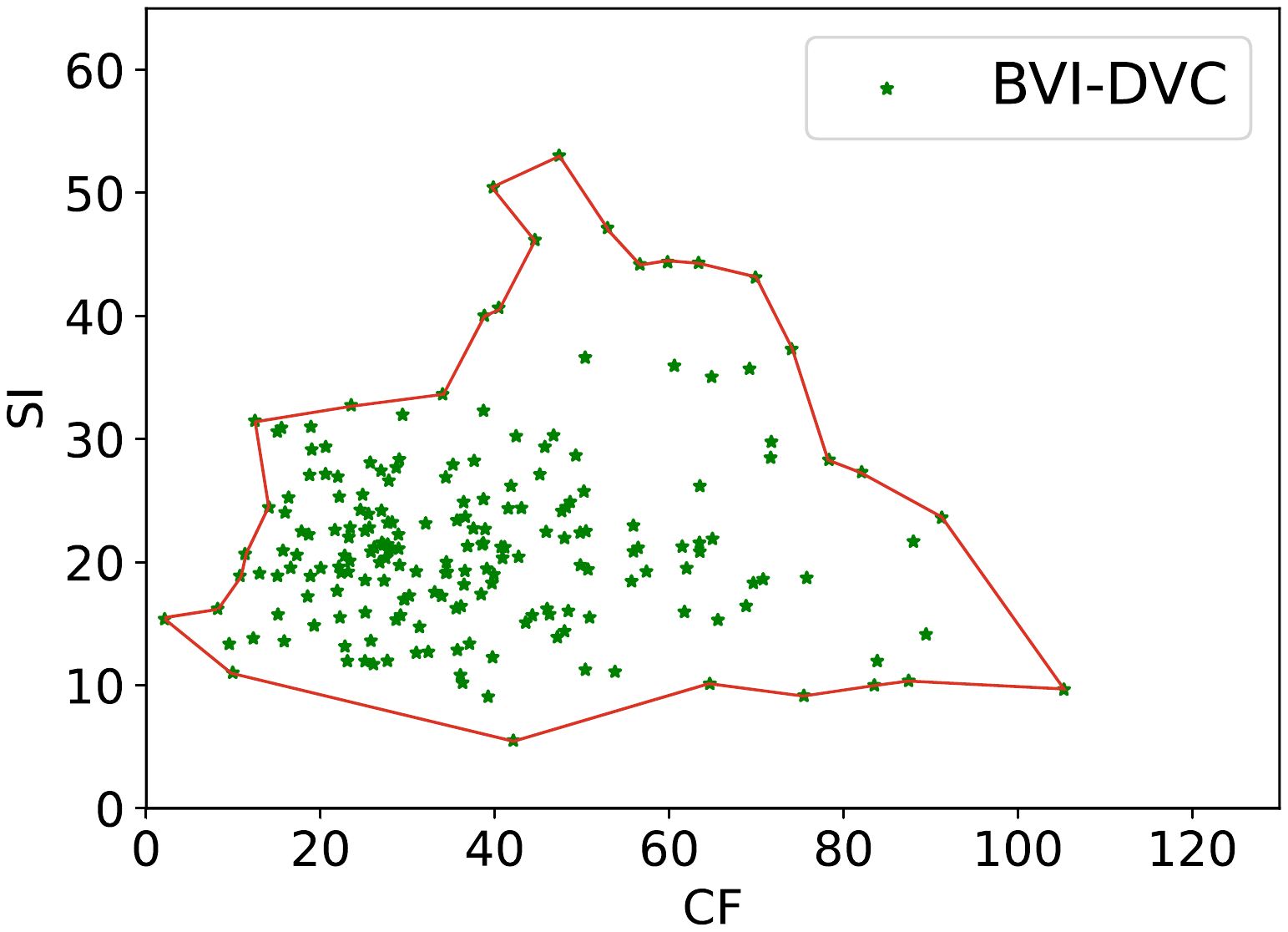}}
    \end{minipage}
    
    \begin{minipage}[b]{0.48\linewidth}
        \centering
        \centerline{\includegraphics[trim={1.9cm 0.5cm 1.8cm 0.5cm},clip,width=\linewidth]{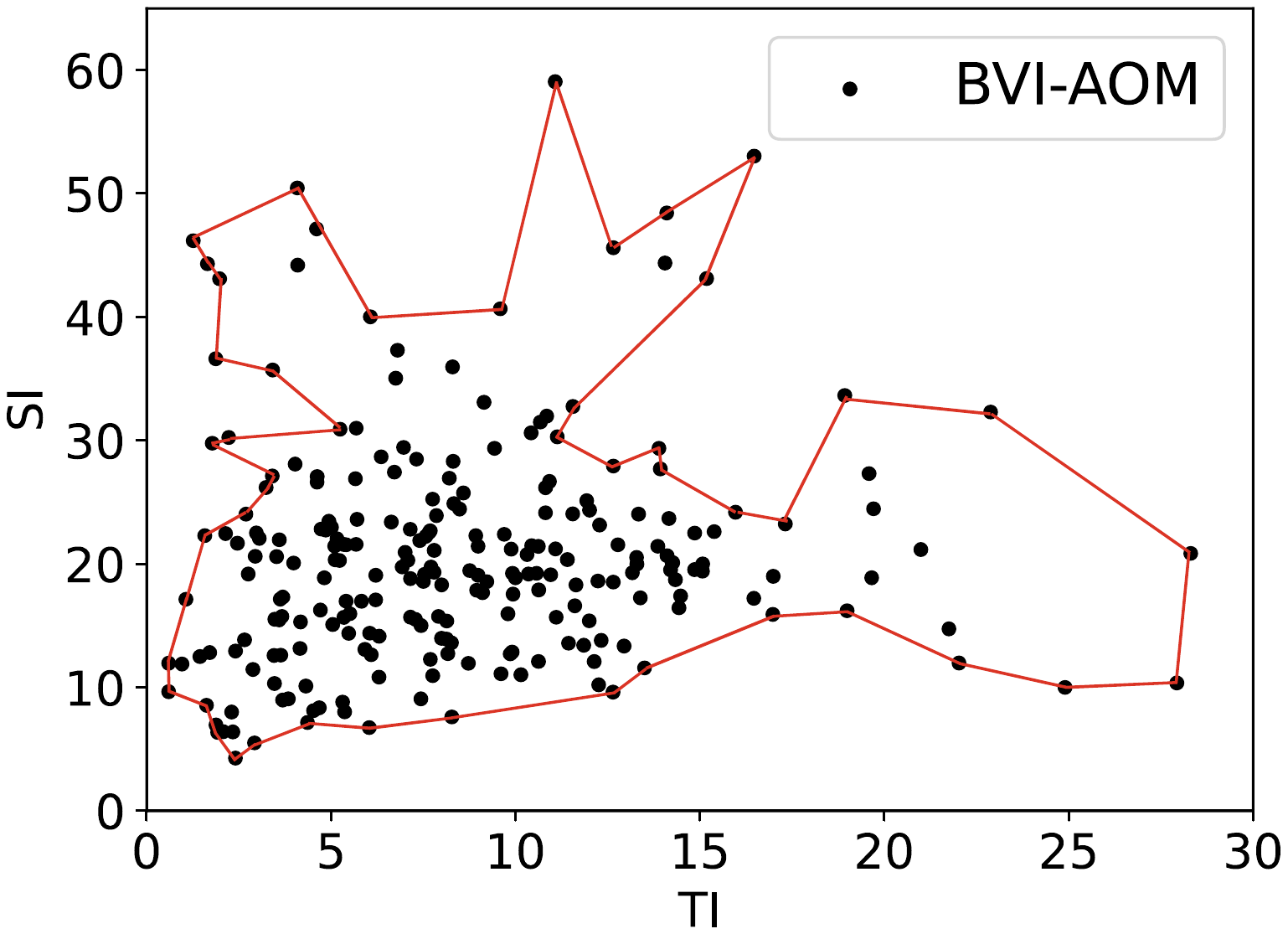}}
    \end{minipage}
    \begin{minipage}[b]{0.48\linewidth}
        \centering
        \centerline{\includegraphics[trim={1.9cm 0.5cm 1.8cm 0.5cm},clip,width=\linewidth]{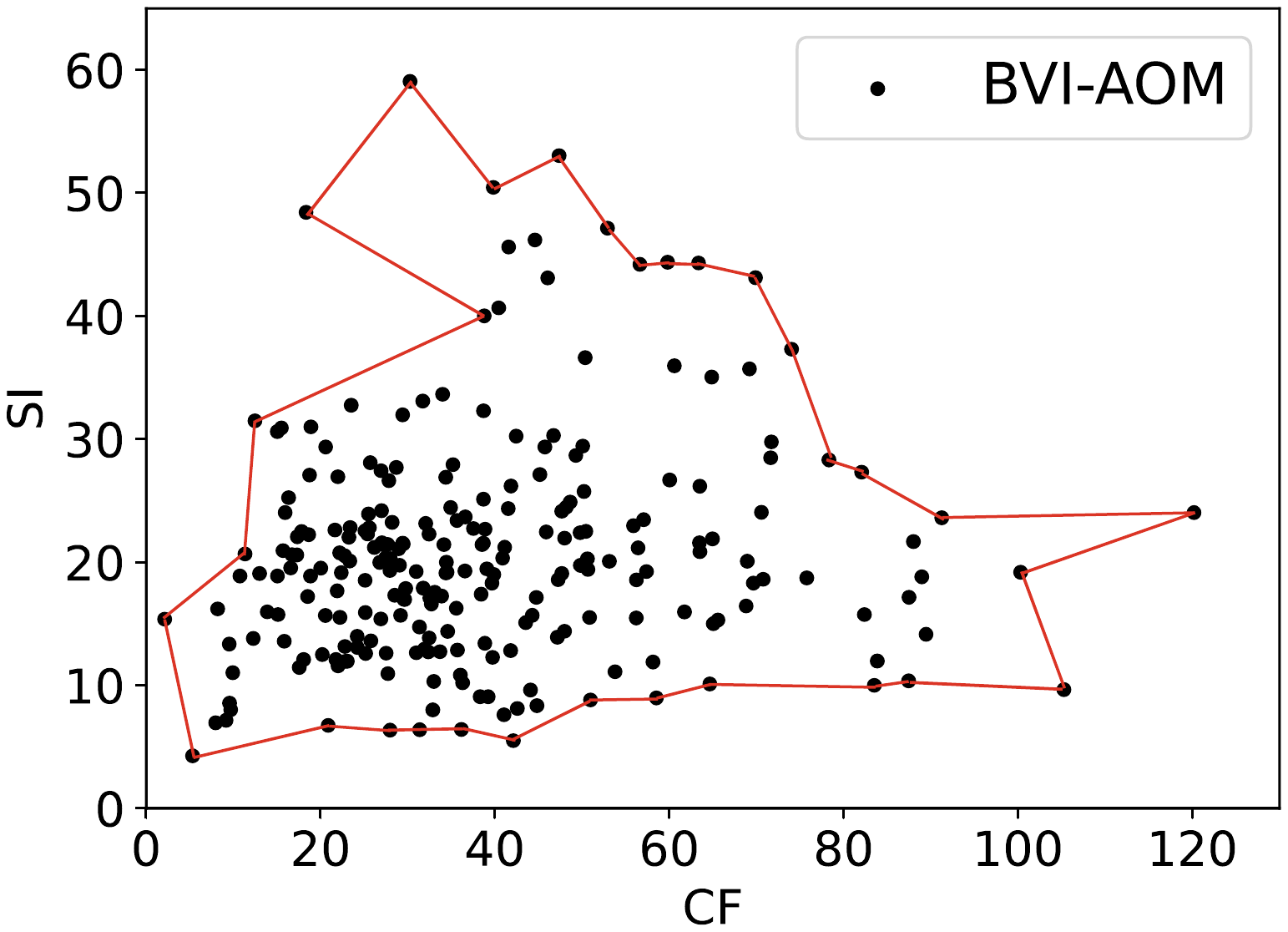}}
    \end{minipage}
    \caption{Distribution of 4K source sequences in the BVI-DVC dataset (upper row) and the BVI-AOM dataset (lower row) in terms of Temporal Information, Spatial Information and Colorfulness.}
    \label{fig:si_ti_cf}
    \vspace{-0.5cm}
\end{figure}

To better support the research on deep video compression, this paper proposes a new training dataset based on BVI-DVC, which is free of content with restrictive licensing terms and offers improved performance when used as a training set for deep video coding solutions. Specifically, this dataset, named BVI-AOM, consists of 956 uncompressed pristine video clips, each 64 frames long, with a spatial resolution from 270p to 2160p. To demonstrate its superior training performance for deep video compression, this dataset has been used to train two popular network architectures for two coding tools integrated with the AOM Video Model (AVM). The results have been compared with those based on the BVI-DVC dataset showing additional coding gains of up to 2.98 percentage points (p.p.). We hope that this new dataset will benefit the video coding community thanks to its flexible copyright license and excellent training performance. 

The remainder of this paper is structured as follows. Section~\ref{sec:db} describes the proposed dataset and quantifies its content coverage. Section \ref{sec:methodology} showcases the design of the experiment aimed at benchmarking the performance of the dataset. The results of this experiment are then summarized and discussed in Section~\ref{sec:results}. Finally, Section~\ref{sec:conclusion} concludes the paper and outlines future work.

\section{BVI-AOM Dataset}
\label{sec:db}

\subsection{Video Sequences}

We follow the same content selection method as in \cite{ma2022bvidvc} and select 239 pristine UHD sequences from the following sources: i) the American Society of Cinematographers Standard Evaluation Material 2 (ASC StEM 2) \cite{stem2}, ii) SVT Open Content Video Test Suite 2022 (SVT2022) \cite{andersson2022svt}, iii) CableLabs 4K sequences \cite{cablelabs}, and iv) the BVI-DVC dataset \cite{ma2022bvidvc}. All of these candidate sequences are in a YCbCr 4:2:0 format with a 10-bit depth. We set the spatial resolution of source sequences to a multiple of 16 (3840$\times$2176 rather than 3840$\times$2160), to make sure that the content can be effortlessly compressed by legacy video codecs which require the resolution of their input to be divisible by 16 both horizontally and vertically. Based on these sequences, we further downsample them to three lower resolutions, 1920$\times$1088, 960$\times$544 and 480$\times$272, using the Lanczos-3 filter implemented in the AVM GitLab repository \cite{avm-repo}. This results in a total of 956 sequences. The technical properties of the BVI-AOM dataset are summarized in TABLE~\ref{tab:properties}. Additionally, Fig.~\ref{fig:thumbnails} shows thumbnails from a set of 15 representative BVI-AOM sequences. Please note that the dataset contains sequences that present not only complex structures (e.g., fire, water, or plasma) but also artistic intent (e.g., action movie like face close-ups).

\begin{table}[t]
    \centering
    \renewcommand{\arraystretch}{1.15}
    \caption{Technical details of the BVI-AOM dataset.}
    \begin{tabular}{r|l}
    \toprule
    Property & Value \\ \midrule
    Pixel format & Planar YUV 4:2:0 \\
    Resolution & 3840$\times$2176, 1920$\times$1088, 960$\times$544, 480$\times$272 \\
    Dynamic range & Standard Dynamic Range (SDR) \\
    Color space & Compliant with Rec. ITU-R BT.709 \\
    Bit depth & 10 bit \\
    FPS & 24 to 120 \\
    No. of frames per seq. & 64 \\
    No. of source seq. & 239 \\
    Total no. of seq. & 239 x 4 {[}resolutions{]} = 956 \\ \bottomrule
    \end{tabular}
    \label{tab:properties}
\vspace{-0.5cm}
\end{table}

\begin{figure*}
        \scriptsize
    \begin{minipage}[b]{0.192\linewidth}
        \centering
        \centerline{\includegraphics[width=.98\linewidth]{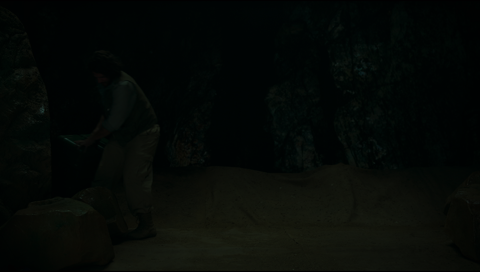}}
        \textcolor[rgb]{0.11,0.46,0.70}{$\bullet$AAscStem2S3}\vspace{.1cm}
        \end{minipage}
    \begin{minipage}[b]{0.192\linewidth}
        \centering
        \centerline{\includegraphics[width=.98\linewidth]{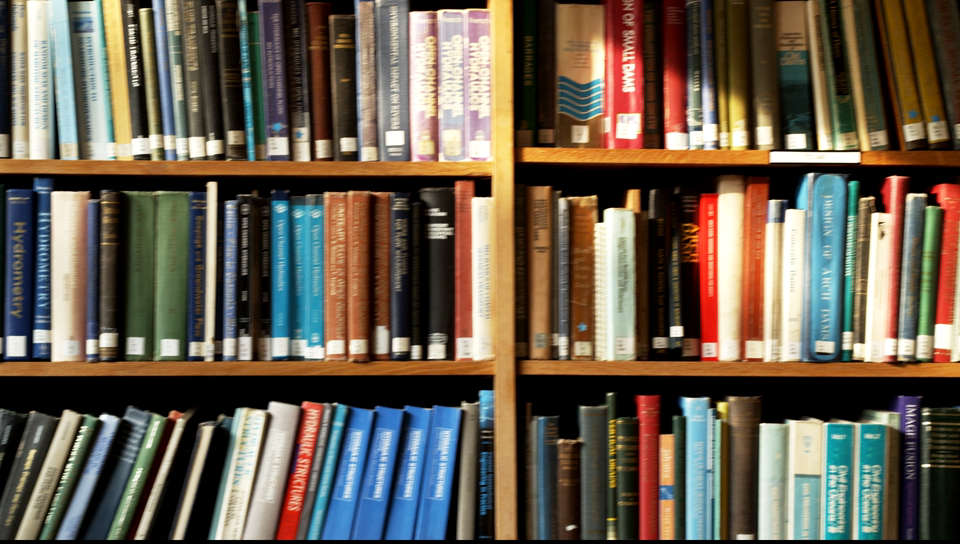}}
        \textcolor[rgb]{1,0.5,0.05}{$\bullet$ABookcaseBVITexture}\vspace{.1cm}
            \end{minipage}
    \begin{minipage}[b]{0.192\linewidth}
        \centering
        \centerline{\includegraphics[width=.98\linewidth]{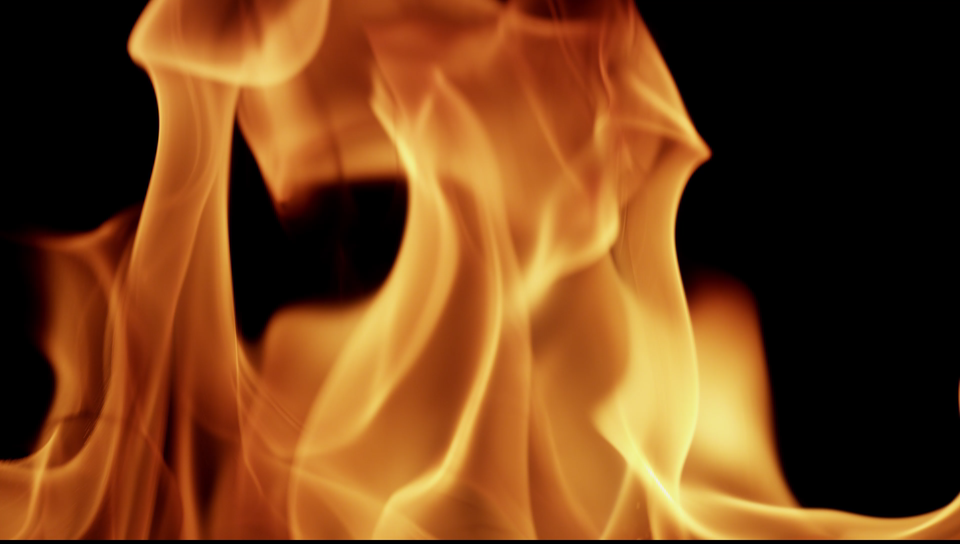}}
        \textcolor[rgb]{0.17,0.63,0.17}{$\bullet$AFireS21Mitch}\vspace{.1cm}
            \end{minipage}
    \centering
    \begin{minipage}[b]{0.192\linewidth}
        \centering
        \centerline{\includegraphics[width=.98\linewidth]{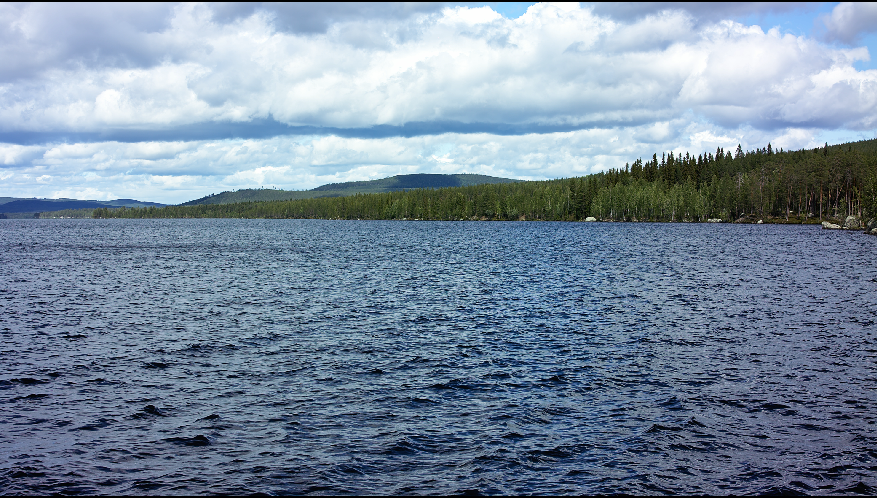}}
        \textcolor[rgb]{0.84,0.15,0.15}{$\bullet$AForestLake}\vspace{.1cm}
            \end{minipage}
    \begin{minipage}[b]{0.192\linewidth}
        \centering
        \centerline{\includegraphics[width=.98\linewidth]{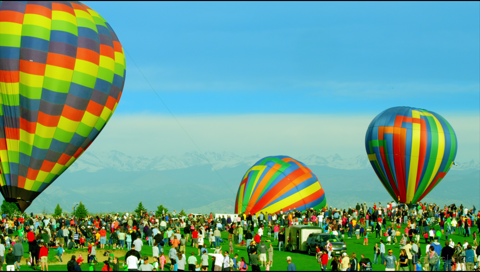}}
        \textcolor[rgb]{0.58,0.40,0.74}{$\bullet$ALiftingOffS7}\vspace{.1cm}
    \end{minipage}
    
    \begin{minipage}[b]{1\linewidth}
        \centering
        \includegraphics[width=\linewidth,clip,viewport=0in 0.2in 14in 3.7in]{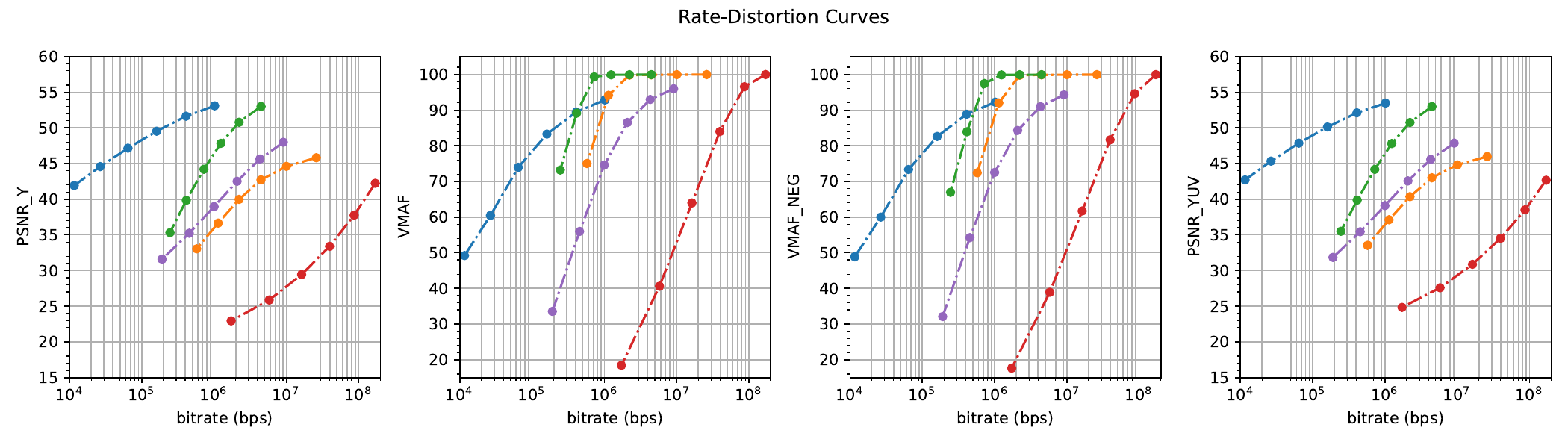}
    \end{minipage}
    \vspace{-0.5cm}
    \caption{Rate-distortion curves for selected outlying BVI-AOM sequences along with the sequence thumbnails.}
    \label{fig:rate_distortion}
    \vspace{-0.5cm}
\end{figure*}

\subsection{Content Coverage}
\label{ssec:content_coverage}
Fig.~\ref{fig:si_ti_cf} shows scatter plots of three video features computed for 239 UHD source sequences contained in the BVI-AOM dataset. The three features are Spatial Information (SI), Temporal Information (TI), and Colourfulness. The calculation of these features is defined in \cite{winkler2012analysis,zhang2018bvi}. To ensure results reproducibility, we follow the recommendation of the Video Quality Experts Group and use SI \& TI implementation from the \emph{siti-tools} GitHub repository \cite{siti-tools}. For CF, we follow the implementation given in \cite{winkler2012analysis}. For comparison, in Fig.~\ref{fig:si_ti_cf} we also show the same plots for the BVI-DVC dataset. It is clear that BVI-AOM achieves better content coverage and diversity compared to BVI-DVC. 

To further quantify how versatile the BVI-AOM dataset is, we selected five outlying sequences and compressed them with the AVM codec (ver. \textit{research-v3.1.0}) using the RA configuration (in accordance with the AOM CTC v3.0 document \cite{aom_ctc_v3}). Our selection criteria for classifying a sequence as outlying were extreme SI, TI, or CF values. Having compressed the sequences, we used four video quality metrics (PSNR-Y, PSNR-YUV, VMAF, and VMAF-neg) to quantify the quality gap between the resultant encodings and the corresponding pristine source sequences, following the recommendation in the AOM CTC. The compression results, shown in Fig.~\ref{fig:rate_distortion} demonstrate that the encoding bitrates span the range from 10 Kbps to 200 Mbps and all the corresponding indices of four video quality metrics also cover a wide range of qualities. These facts further demonstrate the excellent content diversity of the BVI-AOM dataset.

\section{Experiments}
\label{sec:methodology}

\begin{table*}[t]
    \centering
    \caption{BD-rate coding gain over the anchor for the models trained as post-processing and super-resolution tools}
    \begin{tabular}{r|l|cccc|cccc}
    \toprule
     & Model & \multicolumn{4}{c|}{EDSR} & \multicolumn{4}{c}{SwinIR} \\
     \cmidrule{1-10}
     Tools & Dataset & PSNR-Y & PSNR-YUV & VMAF & nVMAF & PSNR-Y & PSNR-YUV & VMAF & nVMAF \\ \midrule
     \multirow{3}{*}{PP} & BVI-DVC & -2.66\% & -2.96\% & -5.36\% & -4.50\% & -2.75\% & -3.35\% & -3.21\% & -3.03\%  \\
     & BVI-AOM & -2.69\% & -3.16\% & -7.48\% & -5.70\% & -3.04\% & -3.78\% & -4.43\% & -3.79\% \\ \cmidrule{2-10}
     & Gain($\uparrow$) [p.p.] & 0.03 & 0.20 & 2.12 & 1.20 & 0.29 & 0.43 & 1.22 & 0.76 \\ \midrule \midrule
     \multirow{3}{*}{SR} & BVI-DVC & -3.87\% & -4.28\% & -8.65\% & -8.40\% & -3.94\% & -5.01\% & -9.51\% & -9.03\% \\
     & BVI-AOM & -4.02\% & -4.49\% & -10.06\% & -9.95\%  & -4.15\% & -5.49\% & -12.49\% & -11.09\% \\ \cmidrule{2-10}
     & Gain($\uparrow$) [p.p.] & 0.15 & 0.21 & 1.41 & 1.55 & 0.21 & 0.48 & 2.98 & 2.06 \\ 
    \bottomrule
    \end{tabular}
    \label{tab:bdrate}
    \vspace{-0.5cm}
\end{table*}

To demonstrate the advantage of using the proposed dataset as a training set for DVC solutions, we have employed it to train two model architectures: i) EDSR (baseline) \cite{lim2017enhanced} and ii) SwinIR (lightweight) \cite{liang2021swinir} for two video coding tools: i) post-processing (PP) \cite{zhang2021video} and ii) super-resolution (SR) \cite{ma2020cvegan}. EDSR and SwinIR are two popular network architectures which have been used in many deep learning based image/video coding tools \cite{ma2022bvidvc,tong2024swin}, while PP and SR were selected here due to their superior performance over standard video codecs compared to other tools and many end-to-end learned video codecs. Furthermore, both PP \& SR tools have also been proposed to be integrated into future coding standards within MPEG and AOM \cite{JVET-AG0071,joshi2023switchable,kim2017dynamic}. Our experimental setup also aligns with that in \cite{ma2022bvidvc}. In addition, for both networks, we followed the training methodology given in their original literature.

When training the models, we have followed the guidelines set out in the AOM CTC v3.0 document and used the AVM codec version \textit{research-v3.1.0} as our baseline. We chose to use the Random Access (RA) configuration of the codec to encode the training data at six QP levels recommended by the AOM CTC document (110, 135, 160, 185, 210, and 235) to generate training content. This resulted in six models (one per QP level) for each network/tool. For SR, before encoding, the input video frames are first downsampled by a factor of two using the Lanczos 3 filter, following the practice in \cite{chen2020overview,joshi2023switchable}. We trained each model using $5\,000$ training batches from each QP group. One training batch consisted of 16 pairs of patches, each of which includes one 96$\times$96 px patch from a pristine source sequence and one 96$\times$96 px patch from a corresponding encoded (and up-sampled, 
 for SR, using the nearest neighbor filter \cite{ma2022bvidvc}) version of that source sequence.

To evaluate the performance of each model, we first encode 48 AOM CTC sequences (Class A) using the same baseline codec and then apply each model to the reconstructed sequences (using a model appropriate for a QP level of a given sequence). For SR, we only perform encoding for eight UHD (Class A1) sequences, as for lower-resolution clips, the coding gains of using SR have been shown to be limited \cite{ma2020mfrnet}. We then compare the rate quality performance to that of the baseline codec (w/o PP or SR, a.k.a. the anchor), and calculate the performance difference using the Bj\o ntegaard Delta (BD) rate metric \cite{bjontegaard2001calculation}. Here, video quality is measured by four different video quality metrics, including PSNR-Y, PSNR-YUV, VMAF, and VMAF-neg, using the \textit{libvmaf} software package (ver. 2.3.1) \cite{libvmaf-repo}. 

To benchmark the proposed database, we then repeated the same experiment for the BVI-DVC dataset \cite{ma2022bvidvc}, which has been used in MPEG for training neural network based coding tools and has been reported to offer improved training performance over other existing training datasets. 

\section{Results and Discussion}
\label{sec:results}

TABLE~\ref{tab:bdrate} summarizes the BD-rate gains over the anchor for the two network architectures and two coding tools when BVI-DVC and BVI-AOM datasets are employed as the training set. From these results, we can clearly see that training with the BVI-AOM dataset resulted in better performance for all network and coding tool combinations. Specifically, for post-processing, training with the BVI-AOM dataset consistently resulted in additional BD-rate gains over the anchor when PSNR-Y, PSNR-YUV, VMAF and VMAF-neg were used for quality assessment, with an average BD-rate gain improvement of 0.78p.p. (percentage points). Similarly, for SR, the additional benefit of using the BVI-AOM dataset to train the models spans the range from 0.15p.p. for PSNR-Y (EDSR) to 2.98p.p. for VMAF (SwinIR). On average, using the BVI-AOM dataset to train an SR tool leads to an additional 1.13p.p. BD-rate gain (on top of that when training with BVI-DVC).

\begin{figure}[!h]
    \centering
    \begin{minipage}{0.485\linewidth}
        \includegraphics[trim={0.5cm 0.6cm .6cm 0.5cm},clip,width=.9\linewidth]{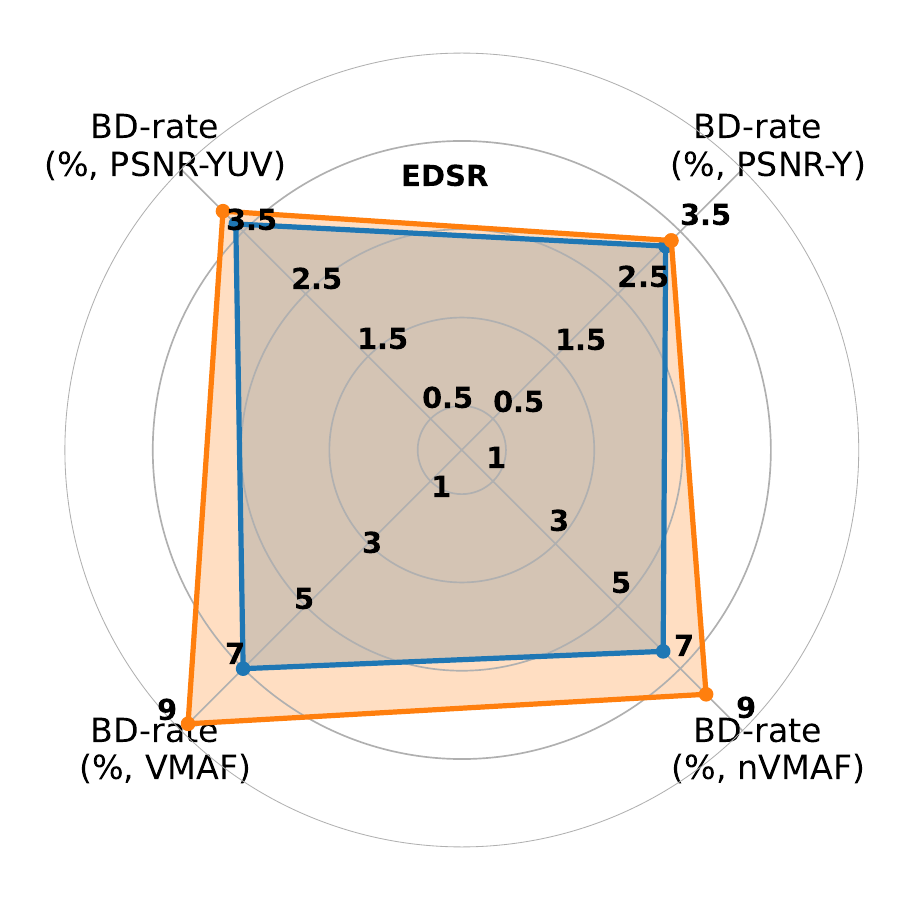}
    \end{minipage}
     \begin{minipage}{0.485\linewidth}
        \includegraphics[trim={0.5cm 0.6cm .6cm 0.5cm},clip,width=.9\linewidth]{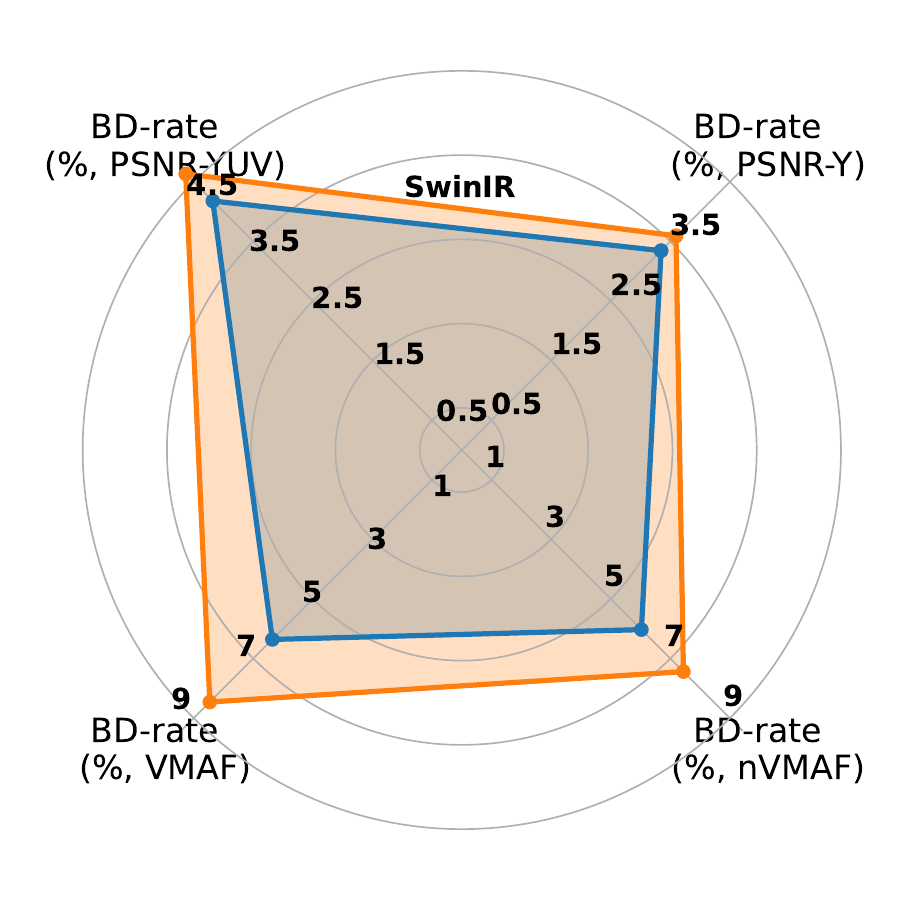}
    \end{minipage}
    
       \begin{minipage}{0.485\linewidth}
        \includegraphics[trim={0.5cm 0.6cm .6cm 0.5cm},clip,width=.9\linewidth]{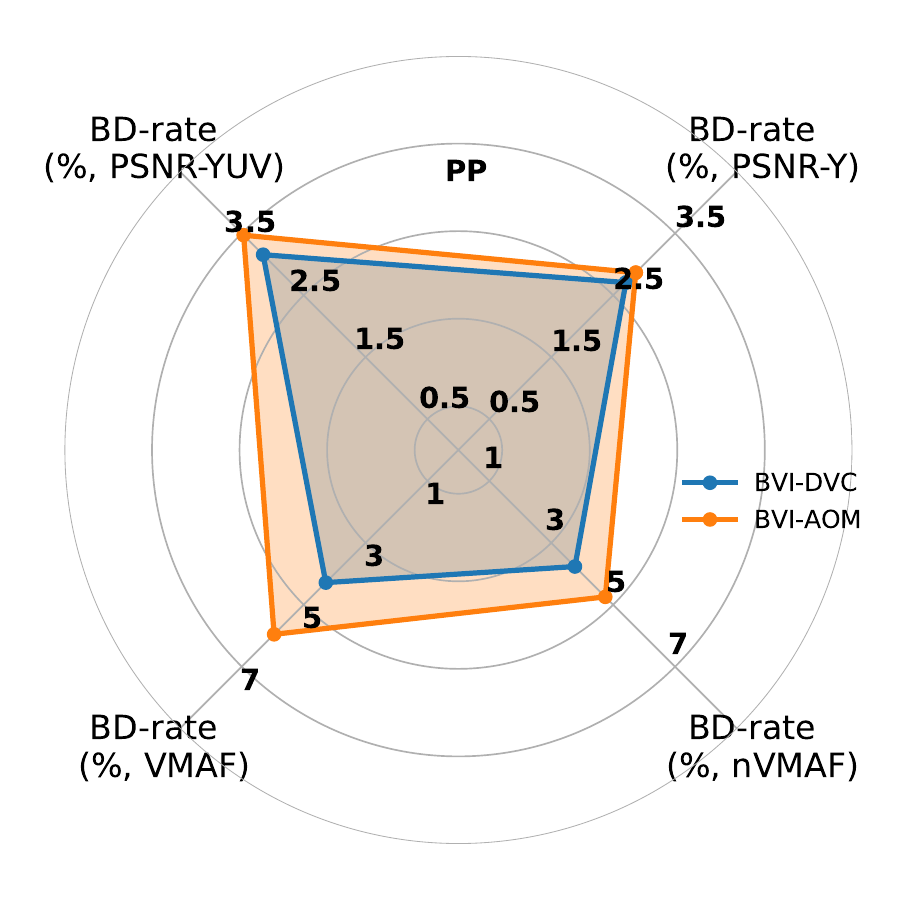}
    \end{minipage}
       \begin{minipage}{0.485\linewidth}
        \includegraphics[trim={0.5cm 0.6cm .6cm 0.5cm},clip,width=.9\linewidth]{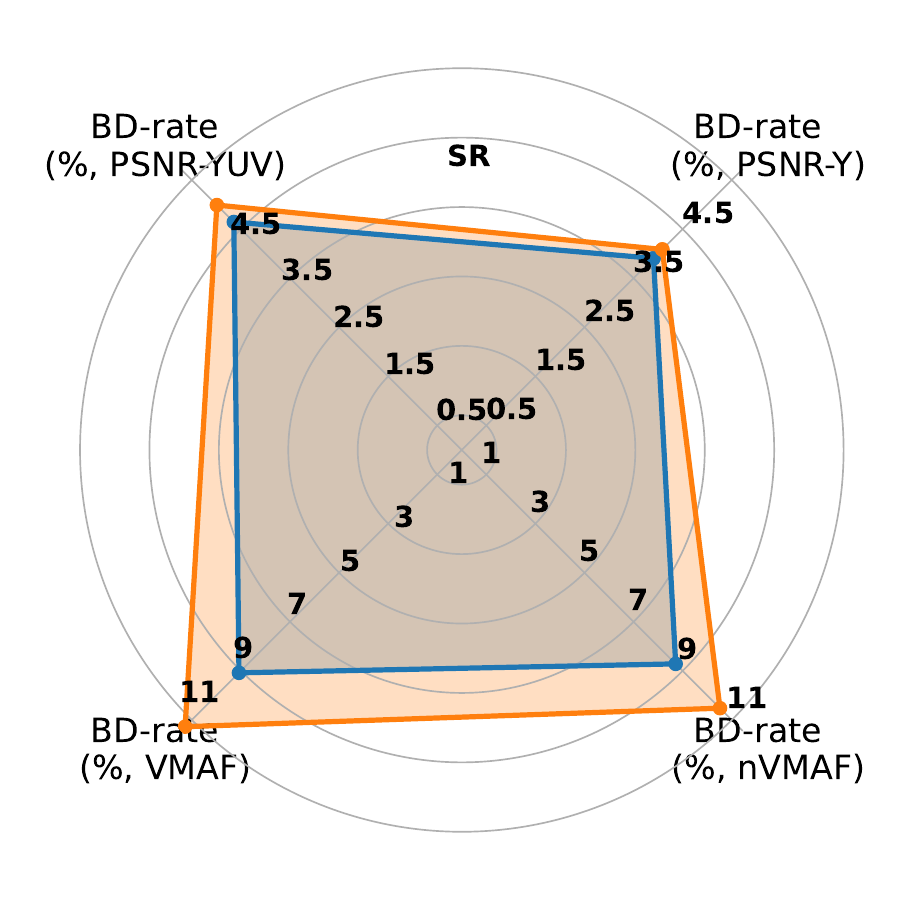}
    \end{minipage}
      
    \caption{Radar charts comparing the average BD-rate savings when models/coding tools are trained with either the BVI-DVC or the BVI-AOM dataset: i) EDSR (for both PP and SR, upper left), ii) SwinIR (for both PP and SR, upper right), iii) PP (for both networks, lower left), and iv) SR (for both networks, lower right).}
    \label{fig:avg_bdrate_dataset}
    \vspace{-0.3cm}
\end{figure}

The superior performance of the proposed dataset is also illustrated in Fig.~\ref{fig:radar-chart} and~\ref{fig:avg_bdrate_dataset}, in which radar graphs are plotted to show the coding performance (in terms of four quality metrics) when BVI-AOM and BVI-DVC are used to train different network structures (EDSR and SwinIR) and coding tools (PP and SR). These also demonstrate that the additional BD-rate gains achieved by BVI-AOM are evident and consistent.

In addition to the improved model generalization and coding performance, the proposed dataset is also associated with more flexible copyright terms compared to BVI-DVC. This will enable it to be used in a wider community and for more diverse applications.

\section{Conclusion}
\label{sec:conclusion}

In this paper, we present a new training dataset, BVI-AOM, for training deep video coding methods. It contains 956 uncompressed sequences with various spatial and temporal resolutions and offers a relatively wide coverage of low-level video features and texture types. When used for training various deep video coding tools, the results show that the BVI-AOM dataset offers consistent performance gains when compared to the commonly used dataset, BVI-DVC, with up to 2.98p.p. additional BD-rate saving. The proposed dataset comes with flexible licensing terms permitting its use for academic research and video standards development purposes. The dataset is publicly available under this link: \url{https://github.com/fan-aaron-zhang/bvi-aom}.

\newpage
\small
\bibliographystyle{ieeetr}
\bibliography{main}

\end{document}